\documentclass[reprint,flushbottom,showpacs]{revtex4-1}
\usepackage{epsfig}
\usepackage{amsmath}
\usepackage{amssymb}
\usepackage{amsthm} 
\usepackage{bm}
\usepackage{color}

\newcommand{\aosc}{a_\mathrm{osc}}
\newcommand{\ket}[1]{\left|#1\right>}
\newcommand{\bra}[1]{\left<#1\right|}
\newcommand{\nn}{\nonumber\\}

\newcommand{\bea}{\begin{eqnarray}}
\newcommand{\ea}{\end{eqnarray}}
\newcommand{\eea}{\end{eqnarray}}
\newcommand{\ord}{{\cal O}}
\newcommand{\sumint}[1]

\begin{document}

\title{Interacting trapped bosons yield fragmented condensate states 
in low dimensions}

\author{Uwe R. Fischer$^{1,3}$ and Philipp Bader$^{2,3}$}

\affiliation{$^1$Seoul National University,   Department of Physics and Astronomy \\  Center for Theoretical Physics, 
151-747 Seoul, Korea\\
$^2$Universidad Polit\'ecnica de Valencia, Instituto de Matem\'atica Multidisciplinar,  
E-46022 Valencia, Spain \\
$^3$Eberhard-Karls-Universit\"at T\"ubingen,
Institut f\"ur Theoretische Physik \\
Auf der Morgenstelle 14, D-72076 T\"ubingen, Germany}

\begin{abstract}
We investigate the 
level population statistics and degree of coherence  
encoded in the single-particle density matrix 
of harmonically trapped low-dimensional [quasi-one-dimensional (quasi-1D) or quasi-two-dimensional (quasi-2D)] Bose gases with repulsive contact interactions.
Using a variational analysis, we derive fragmentation of the condensate in the weakly confining directions into two (quasi-1D) respectively three (quasi-2D) mutually incoherent macroscopic pieces, upon increasing a dimensionless interaction measure beyond a critical value. 
Fragmented condensate many-body states in low-dimensional systems 
therefore occur well before the thermodynamic limit of infinite extension is reached, in which phase fluctuations of the matter wave field create an infinite number of nonmacroscopic fragments.
\end{abstract}


\maketitle

\section{Introduction}
Bose-Einstein condensation, the macroscopic occupation of 
one field operator mode ($\equiv$ single-particle orbital), 
ceases to exist in the thermodynamic limit of infinite extension 
in spatial dimension less or equal to two \cite{Hohenberg,MerminWagner}, as a direct
consequence of the Bogoliubov inequality \cite{Bogoliubov}. 
This so-called Hohenberg-Mermin-Wagner theorem has been argued also to hold for systems 
with a nonvanishing cross-section (quasi-1D) or a finite thickness (quasi-2D) \cite{ChesterFisherMermin}. On the other hand, 
for a finite extension respectively 
radius of curvature along the cylinder axis or in the plane, 
condensation can persist \cite{Fischer}.
Low-dimensional Bose gases are nowadays routinely created by varying cloud aspect ratios in strongly 
anisotropic trapping geometries of magneto-optical or purely optical origin \cite{Goerlitz,Paredes,Hadzibabic}. 
While the very existence of condensates does not depend on interaction in the thermodynamic limit, the 
increasing dependence of their detailed dynamical and static properties on the interaction coupling when lowering 
the dimensionality has been demonstrated experimentally \cite{Kinoshita,Krueger}
as well as investigated theoretically 
\cite{Castin,Lieb}. 

The search for fragmented condensate 
states of Bose gases, that is the macroscopic occupation of more than one field operator mode with no  mutual coherence between them, has a long history, see, e.g., \cite{Penrose,Nozieres,Mueller}. The obvious general difficulty in answering the rather delicate 
question whether a given system is coherent (a condensate) or fragmented consists 
in solving an interacting many-body problem, from which the 
single-particle density matrix follows. The latter, by definition \cite{Penrose}, gives 
the degree of fragmentation versus that of any remaining coherence. 

In the following, we shall show that fragmentation of condensates 
takes place well before the thermodynamic limit is reached in 
lower-dimensional Bose gases. 
We thus argue that interaction-induced fragmented phases, with only a few macroscopically
occupied modes (termed fragmented condensates \cite{Leggett}), 
exist before fragmentation into infinitely many incoherent pieces 
of nonmacroscopic size sets in, occurring when in the thermodynamic limit phase fluctuations destroy 
the condensate in the quasi-1D or quasi-2D low-dimensional regimes \cite{Hohenberg,MerminWagner}.

We concentrate on the point where fragmentation sets in first, into two
(quasi-1D) respectively three (quasi-2D trapping geometry) 
modes with no coherence between them, which may be described by the solution of a many-body problem essentially exactly solvable for any strength of interaction. Thus we demonstrate how the 
formalism laid down in \cite{Bader} leads to fragmentation of a scalar Bose gas in the weakly confining directions. 
It is stressed that for a trapped system positive interaction couplings lead to fragmentation above a critical value, counter to the expectation from the nontrapped thermodynamic continuum limit, where the physics is homogeneous  
and the Fock exchange term in the total energy 
is preventing fragmentation \cite{Nozieres}.

\section{Quasi-1D trapping geometry}  
We begin with the simple case of one excitation direction along the 
weakly confining ($z$) axis of a Bose gas in the quasi-1D limit.  
Assuming that only two (the energetically lowest) longitudinal 
modes are significantly occupied [that is, with macroscopic occupation numbers of $\ord(N)]$, 
we restrict the expansion 
of the field operator to these two modes,  $\hat \Psi ({\bm r}) = \sum_{i=0,1} \hat a_i \Psi_i({\bm r})$; the modes are normalized to unity, $\int d^3r |\Psi_i({\bm r})|^2 =1$.

The validity of the two-mode approximation 
on the mean-field level
and in 
double-well traps has been discussed in 
 \cite{Ostrovskaya,Milburn}. In the following, we go beyond mean-field (solve for the full quantum solution) and consider a single trap, determining the interaction-dependent orbital parameters self-consistently by looking for the ground state variationally (see below). 
The two-mode approximation can then be expected to represent accurately the 
interacting many-body physics in a quasi-1D trapping setup  for much larger values of the coupling constant than one would anticipate from a mean-field analysis with fixed orbitals \cite{Milburn}. 

Consider the interacting two-mode Hamiltonian 
\begin{multline}
\hat H =\sum_{i=0,1}\epsilon_i\hat a_i^\dagger\hat a_i
+\frac{A_1}2 \hat a^\dagger_0\hat a^\dagger_0 \hat a_0 \hat a_0 
+\frac{A_2}2 \hat a^\dagger_1\hat a^\dagger_1\hat a_1\hat a_1 
\\
+\frac{A_3}{2}\hat a_1^\dagger\hat a^\dagger_1\hat a_0\hat a_0
+\frac{A_3^*}2 \hat a_0^\dagger\hat a^\dagger_0\hat a_1\hat a_1
+\frac{A_4}2 \hat a_1^\dagger \hat a_1 
\hat a_0^\dagger\hat a_0 , \label{Ha}
\end{multline}
with the interaction coefficients $A_1= V_{0000}, A_2 = V_{1111}, A_3 = V_{1100}, A_3^*=V_{0011}$, 
and  $A_4 = V_{0101}+V_{1010}+V_{1001}+V_{0110}$, where the interaction 
matrix elements are given by $V_{ijkl} = g\int d^3r
\Psi^*_i({\bm r})\Psi^*_j({\bm r})
\Psi_k({\bm r})\Psi_l({\bm r}) $. The contact interaction is assumed to be 
repulsive, $g = 4\pi a_s>0$, with $a_s$ the  
$s$-wave scattering length (we put $\hbar=m=1$, where $m$ is the boson mass);  
the single-particle energies $\epsilon_1 \ge \epsilon_0$. 
The pair-exchange  coefficients $A_3$ are real or complex numbers
depending on the choice of modes ($A_1,A_2,A_4$ are always real by definition), 
and in general do not vanish in a spatially confined gas.
The spatial basis dependence of the Hamiltonian stemming from the position dependence of the modes' phase is contained in the values of the $A_3$, 
and is reflected in the correlation functions characterizing the response 
of the system to external perturbations. 
Global phase shifts, $\psi_k({\bm r})\rightarrow \psi_k({\bm r})e^{i\theta_k}$ with $\theta_k$ constants, as is well known, leave the correlation functions invariant.

We perform an expansion of the two-mode many-body wavefunction in a Fock basis, 
\bea
|\Psi\rangle=\sum_{l=0}^N\psi_l |N-l,l\rangle,
\ea 
so that $\psi_l$ is the probability 
amplitude for $l$ particles residing in the excited state \cite{Mueller}. 
The total density then is given by 
$\langle \Psi| \hat \rho ({\bm r}) 
|\Psi\rangle 
= \sum_{l=0}^N |\psi_l|^2\left[ (N-l) |\Psi_0({\bm r})|^2 + l |\Psi_1({\bm r})|^2\right]$. 

Nonvanishing pair coherence, as  defined by the expectation value 
$\frac12 \bra \Psi \hat a_0^\dagger\hat a^\dagger_0\hat a_1\hat a_1 + \hat a_1^\dagger\hat a^\dagger_1
\hat a_0\hat a_0 \ket \Psi  
= {\rm Re} \left[ \sum_{l=0}^N d_l \psi_l^* \psi_{l+2}\right]$, where the pair-exchange coefficient   
$d_l= \sqrt{(l+2)(l+1)(N-l-1)(N-l)}$, is enforced by energy minimization,    
and yields definite a value of $A_3$ and coherent pair oscillations 
between the two modes, with $A_3$ equal to their frequency. 
These coherent pair oscillations are thus in analogy to Josephson
oscillations, which are due to an off-diagonal matrix element of the single-particle density matrix
$\bra \Psi \hat a_0^\dagger\hat a_1\ket \Psi \propto N$, analogously represented by single-particle exchange terms of the form
$- \frac\Omega2 \hat a_0^\dagger\hat a_1 + {\rm h.c.}$ in the Hamiltonian, where $\Omega$
is the Josephson frequency. 

Minimizing the energy with respect to the Fock state amplitudes $\psi_l$, 
the linear system to be solved reads 
\bea
E\psi_l = \frac{A_3}{2}(d_l\psi_{l+2}+d_{l-2}\psi_{l-2}) 
+ c_l\psi_l \,,\label{diffeq}
\ea
where $E=\langle \Psi |\hat H|\Psi\rangle$ is the total energy. 
The diagonal coefficient reads $c_l = \epsilon_0(N-l)+\epsilon_1 l+\frac12A_1(N-l)(N-l-1)
+\frac12A_2 l(l-1)+\frac12A_4 (N-l)l$. The above discrete equation connects the values of 
$\psi_l$ on $l$-``sites'' differing by two, and leads to 
alternating signs within the even and odd 
$l$ sectors of the $\psi_l$ when $A_3>0$, 
sign($\psi_l \psi_{l+2})=-1$. As a result, sign($\psi_l \psi_{l+1})=\pm (-1)^l$, the
$\pm$ sign reflecting the initial condition on the $\psi_l$ 
\cite{boundary}. This, then, leads to the 
destruction of first-order coherence defined by 
$g_1=\frac12 \bra\Psi \hat a_0^\dagger \hat a_1  + \hat a_1^\dagger \hat a_0 \ket\Psi = {\rm Re}\left[ \sum_{l=1}^N \psi_{l-1}^*\psi_l \sqrt{l(N-l+1)}\right]$, 
provided $A_1+A_2+2A_3-A_4>0$ \cite{Bader}. 

We take as the (real) modes the ground and first excited state 
oscillator modes along the cylinder axis
\bea
\psi_0(r,z) & = &  \frac{1}{\sqrt{\pi^{3/2}R_z}l_\perp}
\exp\left[-\frac{z^2}{2R_z^2} -\frac{r^2}{2l_\perp^2}\right]
, \nn 
\psi_1(r,z) & = & \frac {\sqrt2 z}{R_z}  \psi_0(z), 
\ea 
where $R_z$ is a {\em variational parameter} for the orbital delocalization length away from the harmonic oscillator ground state in which $R_z \equiv l_z= \omega_z^{-1/2}$; $\omega_z,\,\omega_\perp $ are the harmonic trapping frequencies in the $z$ 
and radial directions, respectively, where $l_z=\omega_z^{-1/2}
,\,l_\perp=\omega_\perp^{-1/2}$ are the corresponding 
harmonic oscillator lengths. 
The transverse degrees of freedom 
are assumed to be frozen, so that the transverse part of the wave function 
is the harmonic oscillator ground state. 
The above choice for the modes leads to $A_3$ real and positive 
(note that generally $A_4 = 4 A_3$ for real modes and contact interaction). 

The single-particle energies of the motion along $z$, occurring in the Hamiltonian \eqref{Ha} [the single-particle energy in the transverse direction is dynamically irrelevant], are $\epsilon_0=\frac{1}{4} R_z^2  \omega_z^2 + \frac{1}{4R_z^2}$ 
and $\epsilon_1 = 3\epsilon_0$, while the interaction coefficients for these orbitals 
are  $A_1=U_0 = {g}/(R_z l_\perp^2 (2\pi)^{3/2})$, $A_i/A_1=\{1,\frac34,\frac12,2\}$. 
Analytical expressions for the ground-state distribution   
can be obtained when the absolute value of the Fock state amplitudes 
$|\psi(l)|$ is considered as a continuum variable \cite{Spekkens}. 
In the continuum limit we have a Gaussian distribution 
for the absolute value of $\psi(l)$, 
\bea
|\psi(l)| =\frac1{(\pi a_{\rm osc}^2)^{1/4}} \exp\left[-\frac{\left(l -\frac N2 -\mathfrak{S}\right)^2}{2a_{\rm osc}^2}\right],    \label{contlimit}
\ea
where the effective oscillator length $\aosc^2 =N\sqrt{\frac{A_3}{A_1+A_2+2A_3-A_4}}$ and  
the shift from a fully fragmented state centered at $l=\frac N2$ reads $\mathfrak{S} =\frac{N(A_1-A_2)/2 
+\epsilon_0-\epsilon_1}{A_1+A_2+2A_3-A_4} $ \cite{Bader}. 
The degree of fragmentation is in the two-mode model defined by 
$\mathfrak{F} = 1-{|\lambda_0-\lambda_1|}/N$, where $\lambda_{0,1}$ 
are the two eigenvalues of the single-particle density matrix 
$\rho^{(1)}_{\mu\nu}=\langle \hat a^\dagger_\mu \hat a_\nu\rangle$, 
resulting in ($N_i = \langle \hat n_i \rangle =  
\langle\hat a_i^\dagger \hat a_i\rangle$)
\bea
\mathfrak{F} = 1-\sqrt{1-\frac4{N^2} \left(N_0 N_1  
- |\langle \hat a_0^\dagger \hat a_1 \rangle|^2\right)}\,.
\ea
Because in the present case first-order 
coherence $ \langle \hat a_0^\dagger \hat a_1 \rangle$ essentially vanishes \cite{boundary},
the fragmentation measure $\mathfrak F$ is simply determined by the mean occupation numbers of the two states.
In the continuum limit, the degree of fragmentation is obtained to be 
${\mathfrak F} = 1-\frac{|{\mathfrak S}|}{N/2}$. Two-mode fragmentation 
is therefore in this limit consistently obtained when $a_{\rm osc}$ is real as well as  
${\mathfrak S} < N/2$ holds, and is becoming maximal when $ {\mathfrak S} = 0$ \cite{Bader}.

The total energy is in the continuum limit \eqref{contlimit} 
for the two-mode model \eqref{Ha}  
given by 
\begin{align}
E = \frac{N^2}{4} A_3\left(\frac{2}{\aosc^2}-1\right) + c_{N/2} -\frac12
\frac{(c_0-c_N)^2}{N^2}\frac{\aosc^4}{A_3 N^2} . \nn
\label{contlimitE} 
\end{align}
The relative values of the interaction coefficients $A_i$ lead to 
$\aosc^2 = \sqrt{\frac23} N $,  
which results in 
$\frac{{E}}{N A_1} 
 =  \frac{N}{3} + \frac{7}3  NX - \frac{8N}{3} X^2$.
The ratio $X=\epsilon_0/(NA_1) 
= 
\frac1{4G_{1}}\left(\frac{1}{\Lambda}+\Lambda^3\right)$,  
is a measure of the ratio of single particle and interaction energies, where the dimensionless
variational parameter is defined by ${\Lambda}= R_z/l_z$.
The shift reads ${\mathfrak S} = \frac{N}{6}\left(1 - 16X\right)$, and the quantity  
\bea
G_{1} = 
\frac{Ngl_z}{(2\pi)^{3/2}l_\perp^2} 
= \frac{Ng_{\rm 1D} l_z}{\sqrt{2\pi}}  
\label{G1D} 
\ea 
is a dimensionless measure of interaction strength.
We used in the final expression on the right-hand side
that the quasi-1D interaction strength reads 
$g_{\rm 1D} = g/(2\pi l_\perp^2) = 2a_s/l_\perp^2 
$ 
(valid when $a_s \ll l_\perp$, i.e. away from geometric scattering resonances \cite{Olshanii}), with dimension of inverse length. 
Note that $G_{1}$ linearly increases with increasing $l_z$, indicating
that single condensates cease to exist for $l_z$ going to infinity 
for any particle number and at any finite interaction strength, which is in 
accordance with what we would expect from previous studies 
\cite{Hohenberg,MerminWagner,Fischer,Pitaevskii}.

\vspace*{-2em}
\begin{center}
\begin{figure}[thb]
\centering
\includegraphics[width=.47\textwidth]{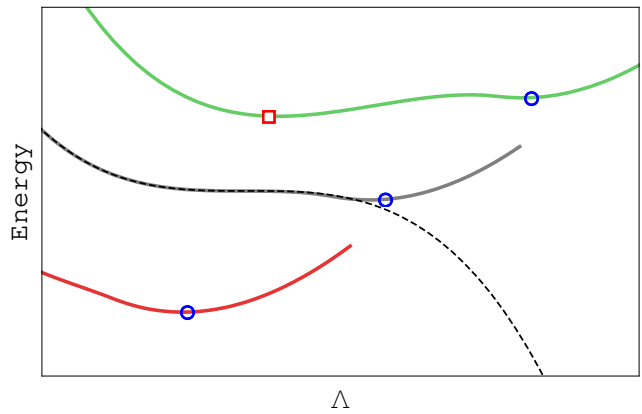}\vspace*{1.25em}
\caption{\label{energyminima} (Color online) 
Evolution (not to scale) of the energy landscape  
as a function of the variational parameter $\Lambda$, with increasing dimensionless interaction measure, 
from bottom to top [the energy minima have additionally 
been shifted relative to each other for clarity]. 
Energy curves are from the full numerical solution of 
\eqref{diffeq}.
The continuum limit is indicated by the dashed curve for the energy curve close to 
criticality (second from the bottom), 
showing that the continuum limit energy, tending to negative infinity for large $\Lambda$,
does not describe the nonfragmented minima of the total energy,  
cf.\,Eq.\,\eqref{ELambdaG}.
Beyond a critical interaction measure, a new minimum corresponding to fragmentation 
appears (empty square), with a discontinous jump from the nonfragmented minimum (circle) 
to a smaller $\Lambda$ (third curve from the bottom). 
Increasing the interaction measure further, the nonfragmented minimum
eventually completely disappears.} 
\end{figure}
\end{center} 

The total energy obtained from solving the discrete many-body equations 
is to be minimized to obtain $\Lambda=\Lambda (G_1)$. 
The analysis reveals that the numerically determined energy as a function of $\Lambda$, for a given $G_1$, has only one, nonfragmented, minimum for small enough $G_1$ below a critical value, while above the latter, a second local minimum appears.  
The transition to a {\em fragmented state}, corresponding to the second minimum at a smaller (scaled) delocalization length $\Lambda$, is discontinous, and the degree of fragmentation jumps to a finite value, cf.\,Fig.\,\ref{energyminima}.  Therefore, the phase transition between single
condensate and fragmented condensate states obtained within the two-mode model \eqref{Ha} is of first order. Increasing the dimensionless 
interaction measure further, the nonfragmented local minimum gradually disappears 
(asymptotically changing into a turning point), and only the fragmented minimum of the total two-mode energy remains (top curve in Fig.\,\ref{energyminima}).  

We obtain from the continuum limit Eq.\,\eqref{contlimitE} for the energy 
per particle in terms of $G_{1}$ and $\Lambda$ the following polynomial equation, using
that $G_1\omega_z/(NA_1) = \Lambda$,  
\begin{align}
	\frac{{E}}{N \omega_z}		 
					 = \frac{7}{12}\left(\frac{1}{\Lambda^2}+\Lambda^2\right)- \frac{1}{6G_{1}}\left(\frac{1}{\Lambda^3} +2\Lambda + \Lambda^5\right) 
					 + \frac{G_{1}}{3 \Lambda}. \label{ELambdaG} 
\end{align}
For large $\Lambda$, we may approximate the above 
equation as
$	\frac{\mathrm{E}}{N\omega_z} \simeq \frac{7\Lambda^2}{12} - \frac{\Lambda^5}{6G_1} + \frac{G_1}{3 \Lambda}$. 
The extrema equation is thus quadratic in $\Lambda^3/G_1$, so that $\Lambda$ is 
scaling with $G_1^{1/3}$ in this limit; 
the minimum solution describing a fragmented state is $\Lambda^3 = \frac{2}{5}G_1$
(the other solution does not describe a minimum, cf.\,Fig.\,\ref{energyminima}, dashed line). 
Note that the large $G_1$ scaling $\Lambda \propto (Na_s l_z/l_\perp^2)^{1/3}$ 
is identical to the single-condensate Thomas-Fermi scaling \cite{Goerlitz} 
[with a different prefactor]; the quantity $X=\epsilon_0/(NA_1)$ asymptotes to  1/10 in this limit.  

Evaluating in the large $G_1$ limit the difference between the fragmented 
energy minimum of \eqref{ELambdaG} at $\Lambda = (2G_1/5)^{1/3}$ 
and the energy obtained from a single condensate 
of all particles residing in the energetically lower mode, 
$E_{\rm gs}= \frac1{4\Lambda^2} + \frac{\Lambda^2}4 + \frac{G_1}{4\Lambda}$, which has a minimum at
$\Lambda = G_1^{1/3}$, we get
\cite{Corro} 
\bea 
\frac{\Delta E}{N\omega_z} = \frac{E_{\rm gs}}{N\omega_z} 
-\frac{E}{N\omega_z} 
\simeq 0.02\times G_1^{2/3}. \label{deltaElargeG1}
\ea
The energy difference between single condensate and fragmented condensate 
ground states is therefore very small around the transition point, increasing with the particle
number like $N^{2/3}$.

\begin{center}
\begin{figure}[!hbt]
\centering
\includegraphics[width=.47\textwidth]{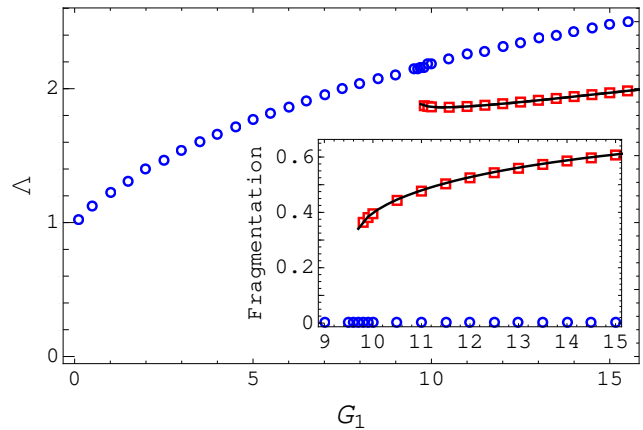}
\caption{\label{FragmentFig1D} (Color online)
Optimal value of the variational parameter $\Lambda$, and degree of fragmentation $\mathfrak F$ (inset), 
both as a function of the dimensionless measure of interaction strength 
$G_1$, in the quasi-1D trapping geometry.
Numerical solutions of the discrete set of many-body  equations \eqref{diffeq} 
are depicted by blue circles (nonfragmented energy minima), and by red squares, respectively, 
the latter describing the energy minima  at which fragmentation occurs, cf.\,Fig.\,\ref{energyminima}. 
The solution in the continuum limit of Eq.\,\eqref{ELambdaG} 
is shown for the fragmented state by the black solid line. The number of particles is $N=500$.}
\end{figure}
\end{center}

We obtain in the continuum limit for the degree of fragmentation
\bea
{\mathfrak F}(G_1) = 1- \frac{1}{3} \left| 1 -  \frac{4}{\Lambda G_1}-\frac{4\Lambda^3}{G_1}\right| 
\simeq \frac45 -\frac{160^{1/3}}{3 G_1^{4/3}},
\nn
\label{Fragment1D}
\ea 
where the last inequality represents the large $G_1$---large $\Lambda$ limit. 
The maximally achievable two-mode degree of fragmentation from this variational ansatz for the modes is thus 80\,\%.
We conclude that fragmentation of the condensate many-body state occurs for 
sufficiently strong interaction coupling at a given finite extent $R_z$ of the cloud, 
that is when 
\bea
G_1 > (G_1)_c \simeq 9.8 
\ea 
where the critical value $(G_1)_c$ for a fragmented minimum of the total energy to first appear,
cf.\,Fig.\ref{energyminima},  
has been determined from the full numerical solution of 
\eqref{diffeq}. 
We display the continuum result for 
$\mathfrak F={\mathfrak F}(G_1)$ obtained from the solution of Eq.\,\eqref{ELambdaG},  
in Fig.\,\ref{FragmentFig1D}. The system is for small interactions and densities ($G_1\ll 1$) 
a single condensate and $\Lambda$ is close to unity; the slow increase of the optimal $\Lambda$ with $G_1$ additionally 
demonstrates that an ansatz incorporating only the two lowest harmonic oscillator 
states is a reasonably accurate approximation. 
For large shift (large $\Lambda$), 
the continuum description fails, 
and the full discrete coupled system of equations for $\psi_l$ in Eq.\,\eqref{diffeq} has to be solved to find the 
optimal value of $\Lambda$. The result from the (exact) two-mode many-body wavefunction corresponding
to the numerical solution of \eqref{diffeq} is displayed together with the continuum limit 
in Fig.\,\ref{FragmentFig1D}, where very good agreement is visible.
\section{Quasi-2D trapping geometry} 
Consider now a cylindrically symmetric trap, with an infinite continuum of 
excitation directions in real space, that is, in all radial directions. A crucial difference to the quasi-1D case 
is the topology, resulting in an angular quantum number $m_\phi$. Furthermore, the lowest ($m_\phi = \pm 1$) 
azimuthal excitation modes of the cylindrically symmetric harmonic oscillator 
are lower in energy than the radial excitations. 
The harmonic oscillator ground state in radial as well as transverse 
directions and the two lowest excited states 
along the azimuthal direction are given by
\bea
\psi_0 (r,z) & = &  \frac1{\sqrt {\pi^{3/2} l_z} R_\perp} \exp\left[-\frac{r^2}{2R_\perp^2} -\frac{z^2}{2l_z^2}\right],
\nn \psi_{\pm} (r,\phi,z) & = & \frac r{R_\perp} \exp[\pm i\phi] \psi_0, \label{threemodes}
\ea 
and are thus the three lowest-lying single-particle states.

\subsection{Two-mode approximation}\label{2msection} 
We first investigate whether fragmentation occurs between the ground state 
and a (coherent) superposition of the left- and right-circulating azimuthal waves 
$\psi_+, \psi_{-}$. We shall consider below the full three-mode problem of taking all three states in \eqref{threemodes} into account, 
and will show that in the continuum limit the corresponding predictions 
become essentially identical to those of the simpler two-mode model, which we discuss here to most clearly elucidate the primary differences to the quasi-1D trapping geometry.

We consider an equal-weight coherent superposition for the excited state 
wavefunction (such that it has angular momentum zero like the 
ground state),  
\begin{multline}
\psi_1  = \frac1{\sqrt2} (\psi_+ + \psi_{-}) \\
= \sqrt{\frac{2}{\pi^{3/2} R_\perp^4 l_z }} \, r \cos\phi\, \exp\left[-\frac{r^2}{2R_\perp^2} -\frac{z^2}{2l_z^2}
\right],  \label{coherentsuperpos}
\end{multline}
and evaluate the coefficients parametrizing the two-mode Hamiltonian \eqref{Ha}.
We obtain $A_1= {g}/{((2\pi)^{3/2} R_\perp^2 l_z)}\equiv V_0 = \frac{G_2 \omega_\perp}{N \Lambda^2}$ and $A_i/A_1=\{1,\frac34,\frac12,2\}$ with 
$\Lambda =R_\perp/l_\perp$;  
the ratios of the interaction coefficients are therefore identical to the quasi-1D case. 
We furthermore have 
$\epsilon_0 = \frac{1}{2R_\perp^2}+ \frac12 \omega_\perp^2 R_\perp^2$ and $\epsilon_1 = 2\epsilon_0 $.
The result for the shift is $\mathfrak S = \frac N6 (1- 8 X)$, with 
$X= \frac{\epsilon_0}{NA_1} 
=\frac{1+\Lambda^4}{2G_{2}}$.    
The dimensionless interaction measure now is defined to be
\bea
G_{2}  = \frac{Ng}{(2\pi)^{3/2} l_z} = \frac{Ng_{\rm 2D}}{2\pi}. 
\label{G2D}
\ea 
It involves $g_{\rm 2D}=g/(\sqrt{2\pi}l_z)$,  
the quasi-2D coupling constant when $l_z \gg a_s$, 
i.e., away from geometric scattering resonances \cite{Holzmann}.
Observe that the dimensionless interaction strength $G_2$ here is {\em independent} of the harmonic oscillator length, in marked distinction to its 1D counterpart $G_1$ in \eqref{G1D}. 
This independence characterizes the 2D trapped case as marginal \cite{Fischer}, in the
sense that for any finite trapping length (nonzero trapping frequency), single condensates persist 
as long as the interaction coupling is sufficiently small. Physically, this
corresponds to the fact that the usual logarithmic divergence of (the integral of) 
phase fluctuations in two spatial dimensions  \cite{Bogoliubov},  is cut off by the trapping.

The calculation then proceeds along the same lines as in the quasi-1D case  and, again 
using \eqref{contlimitE},  we have  
$\frac{{E}}{N A_1} = \frac{N}{3} + \frac{5}3  NX - \frac{2N}{3} X^2$. 
The continuum limit energy expression has a form analogous to Eq.\,\eqref{ELambdaG}, 
using the relation $G_2\omega_\perp/(NA_1) = \Lambda^2$, 
\begin{align}
	\frac{{E}}{N \omega_\perp}		 
					 = \frac{5}{6}\left(\frac{1}{\Lambda^2}+\Lambda^2\right)
					 - \frac{1}{6 G_{2}}\left(\frac{1}{\Lambda^2} +2\Lambda^2 + \Lambda^6\right) 
					 + \frac{G_{2}}{3\Lambda^2} . \label{ELambdaGQUASI2DTwoState} 
\end{align}
Note, however, the different power law dependence on $\Lambda$ in the interaction terms.
\begin{center}
\begin{figure}[t]
\centering
\includegraphics[width=.47\textwidth]{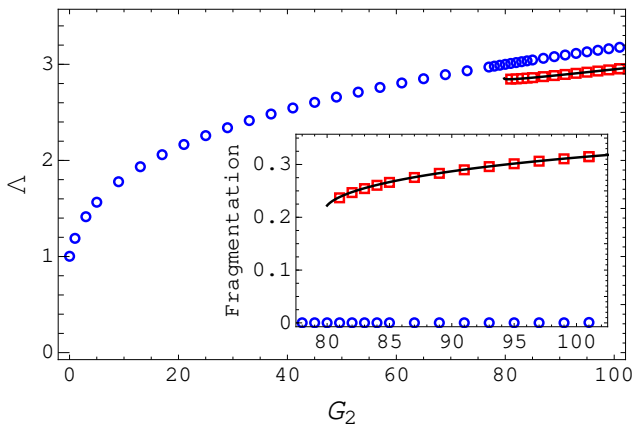}
\caption{\label{FragmentFig2D} (Color online) 
Energy minima of the two-mode ansatz for the quasi-2D trapping geometry 
in terms of the variational parameter $\Lambda$, and degree of fragmentation $\mathfrak F$ (inset), 
both as a function of the interaction measure $G_2$. 
Conventions are the same as those in the quasi-1D trapping setup of 
Fig.\,\ref{FragmentFig1D}. 
Note that the decrease of the size $\Lambda$ 
for fragmented states relative to that of single condensate states 
is significantly less than in the quasi-1D case. The number of particles here is
$N=20000$; a higher particle number is necessary to achieve 
agreement of numerics and analytics comparable to that of the quasi-1D case.} 
\end{figure}
\end{center} 

The minimum equation obtained by differentating \eqref{ELambdaGQUASI2DTwoState} with respect to $\Lambda$ now is solvable analytically; it is a quadratic equation in 
$\Lambda^4$, yielding 
\bea
(\Lambda_{\rm 2m})^4 = \frac16 \left(5G_2-2-\sqrt{16-80 G_2+(G_2)^2}\right) \label{2mLambda}
\ea 
at the fragmented energy minimum.  
The large $G_2$ scaling $\Lambda \propto (Na_s /l_z)^{1/4}$ 
is, like in the quasi-1D setup, identical to the single-condensate 
Thomas-Fermi scaling \cite{Goerlitz}; the quantity $X=\epsilon_0/(NA_1)$ asymptotes here to 1/3.
In the large $G_2$ limit the energy difference between the fragmented 
energy minimum of \eqref{ELambdaGQUASI2DTwoState} and the energy minimum of 
$E_{\rm gs}= \frac12 (1/\Lambda^2 +\Lambda^2)+G_2/(2\Lambda^2)$
[again assuming all particles to be in the energetically lower mode, i.e., 
that the many-body state is a simple condensate or equivalently a $l=0$ Fock state], 
at $\Lambda = G_2^{1/4}$, is given by 
$\frac{\Delta E}{N\omega_\perp} = \frac{E_{\rm gs}}{N\omega_\perp}-\frac{E}{N\omega_\perp} 
\simeq 0.002\times G_2^{1/2}$. 
The energy difference between condensate and fragmented ground states thus
increases more slowly than in the quasi-1D case with the dimensionless interaction measure, cf.\,Eq.\,\eqref{deltaElargeG1}. 

 
The two-mode fragmentation measure gives 
\bea{\mathfrak F}(G_2) = 1- \frac{1}{3} \left|1-\frac{4(1+\Lambda_{\rm 2m}^4)}{G_2}\right|  
\simeq \frac49 - \ord\left(\frac1{G_2}\right).
\label{F2mcont} 
\ea 
The maximally achievable two-mode degree of fragmentation from the ansatz \eqref{coherentsuperpos} is thus 44\,\%. 
Fragmentation takes place when the dimensionless coupling
(cf.\,Fig.\,\ref{FragmentFig2D}, where we compare numerical and continuum limit analytical 
results for the fragmentation) 
\bea 
{G_2} > {(G_2)_c} \simeq 80 . 
\label{G2crit2m} 
\ea
Due to the independence of $G_2= \sqrt{2/\pi} \, Na_s/l_z$ on the trapping in the plane, 
condensate fragmentation in quasi-2D obtains at fixed transverse trapping solely 
because of an increase of the combination $Na_s$.

\subsection{Three-mode approximation} \label{3msection}
We now consider all three modes in \eqref{threemodes}
as the independent field operator modes, $\hat \Psi ({\bm r}) = \sum_{i=0,\pm} \hat a_i \Psi_i({\bm r})$.  
A significant simplification takes place in the Hamiltonian due to the exact vanishing of the pair-exchange coefficients involving two particles of the same mode $\pm$ and the single-particle ground state 
owing to the cylindrical symmetry of the excitation modes. The remaining 
pair-exchange process is between a pair of counter- and copropagating azimuthal
excitations and the ground state. The Hamiltonian then reads 
\begin{multline}
\hat H = \sum_{i=\{0,\pm\}} \epsilon_i \hat n_i  + \frac{B_1}2 \hat n_0 (\hat n_0-1) 
+ \sum_{j=\{\pm\}}\frac{B_2}2 \hat  n_j (\hat n_j - 1) \\
 + \frac{B_3}2 \hat n_0\left[ \hat n_+ + \hat n_{-}\right]  +  \frac{B_4}2 \hat n_+ \hat n_{-} + \frac{B_3}4 \hat a_0^\dagger \hat a_0^\dagger \hat a_+ \hat a_- + {\rm h.c.}
\label{threemodeH}
\end{multline}
The interaction coefficient $B_1 = 2\pi g\int r dr dz \psi_0^4 = 
 {g}/({(2\pi)^{3/2} R_\perp^2 l_z})$, and   
\bea
B_1= \frac{G_2 \omega_\perp}{N \Lambda^2}, \qquad 
{B_i}/{B_1} = \left\{1,\frac12,2,2\right\} . \label{B_i}
\ea
The single-particle energies are $\epsilon_0/\omega_\perp = \frac{1}{2\Lambda^2} + \frac12 \Lambda^2$ 
and $\epsilon_+ =\epsilon_- = 2\epsilon_0$. 
We evaluate the total energy in the three-state Fock basis, with the 
following ansatz for the state vector 
\bea
\ket{\Psi} = \sum_{l,l'} \psi_{l,l'} \ket{N-l-l',l,l'} , \label{3mansatz}
\ea
that is with the probability amplitudes $\psi_{l,l'}$ for 
$l$ and $l'$ particles being in the co- and counterpropagating modes, respectively. Note that the excited state \eqref{coherentsuperpos} in the two-mode approximation 
corresponds in the three-mode basis to setting $\ket{\Psi} = \frac1{\sqrt{2}}
\sum_l\psi_l \left(\ket{N-l,l,0} + \ket{N-l,0,l}\right)$. 

The ansatz \eqref{3mansatz} results in the following functional of the total
energy 
\vspace*{0em}
%
\begin{widetext}
\bea
\langle \Psi |\hat H|\Psi\rangle  & =&  \sum_{l,l'=0}^N  |\psi_{l,l'}|^2 \left[ 
\epsilon_0 N + (\epsilon_1-\epsilon_0)(l+l') 
+\frac{B_1}2 (N-l-l') (N-l-l'-1) 
+\frac{B_2}2 \left[l(l-1)+l'(l'-1)\right] 
\right.\nn 
& & 
\left. 
+\frac{B_3}2 (N-l-l')(l+l') 
+ \frac{B_4}2 l l' \right]
+ \frac{B_3}4 d_{l,l'} \psi^*_{l,l'} \psi_{l+1,l'+1} +{\rm h.c.}
\,,
\ea
\end{widetext}
where the pair-exchange coefficient here reads 
$d_{l,l'} = [(N-l-l'-1)(N-l-l')(l+1)(l'+1)]^{1/2}$. 
We minimize with respect to the state amplitudes $\psi_{l,l'}$,  
to get 
\bea
E \psi_{l,l'} & = & c_{l,l'} \psi_{l,l'} + \frac{B_3}4 d_{l,l'} \psi_{l+1,l'+1} \nn
& & + \frac{B_3}4 d_{l-1,l'-1} \psi_{l-1,l'-1}  \label{FullEV2D}
\ea
where $c_{l,l'} = \epsilon_0 N + (\epsilon_1-\epsilon_0)(l+l') 
+\frac{B_1}2 (N-l-l') (N-l-l'-1) 
+\frac{B_2}2 \left[l(l-1)+l'(l'-1)\right] 
+\frac{B_3}2 (N-l-l')(l+l') 
+ \frac{B_4}2 l l' . 
$ 

We argue that the peculiar matrix structure of \eqref{FullEV2D} allows for an understanding of the $(N+1)(N+2)/2$ dimensional eigenvalue problem in $(l,l')$ in the 
form of $N+1$ smaller problems indexed by $k=0,\ldots,N$ [where $k=l'-l$], 
and only dependent on one running variable $l$.
To see this, we rewrite the state amplitudes $\psi_{l,l'}\equiv \psi_{l,l+k}\equiv \psi_l^k$ and also the matrix elements $c_{l,l'},d_{l,l'}$; then, we note that only diagonal coupling terms occur in \eqref{FullEV2D}, i.e. 
\bea\label{FullEV2Dmod}
E \psi_{l}^k = c_{l}^k \psi_{l}^k + \frac{B_3}4 d_{l}^k \psi_{l+1}^k
+ \frac{B_3}4 d_{l-1}^k \psi_{l-1}^k \nn
\ea
where the superscript $k$ runs from $0$ to $N$, thereby restricting the corresponding $l\in\{0,\ldots,N/2-\left\lceil k/2\right\rceil\}$.
The symmetry of the eigenvalue problem with respect to the exchange of $l$ and $l'$ justifies the reduction to $l\leq l'$ that has been implicitly performed by choosing the ranges of $k$ and $l$. Hence 
all states at a given $k\neq0$ are at least twofold degenerate.
The above reformulation of \eqref{FullEV2D} 
elucidates that different values of $k$ belong to uncoupled equations and thus allows us to study the lower dimensional problems \eqref{FullEV2Dmod} independently.
In matrix language, this observation can be understood as a particular ordering of the $(l,l')$ indices, such that the eigenvalue problem decomposes into a block diagonal matrix,  with each of the $N+1$ blocks corresponding to a symmetric tridiagonal matrix for a given $k$ by Eq.\,\eqref{FullEV2Dmod} and with block sizes $N/2+1-\left\lceil k/2 \right\rceil$,
where the square brackets indicate the next larger integer.




In the large $N$ limit, we again apply the continuum limit \cite{Bader} to \eqref{FullEV2Dmod}, respecting the alternating
signs for even and odd values of $l$ in the $\psi^k_l$, which occur due to the repulsive pair-exchange coupling $B_3>0$, cf. the corresponding distribution in the 
quasi-1D case, Eq.\,\eqref{contlimit}. 
Taking the continuum limit yields
\begin{equation}
	- \frac{1}{2m_0} \Delta |\psi^k| + \frac12 m_0\Omega^2 (l-\mathfrak{S})^2 |\psi^k| = \left(E - E_0\right)|\psi^k|
\end{equation}
with inverse mass coefficient of the harmonic oscillator analogy 
$1/m_0 = \frac{G_2 \omega_\perp}{8N\lambda^2} \sqrt{(N-k)^3 (N+3k)}$,  
the effective frequency
\bea
\Omega & = & \frac{G_2 \omega_\perp}{\sqrt{2}N\Lambda^2}\sqrt{
k^2 + N^2 - \frac{8 k^3}{N+3k} - \frac{\sqrt{(N - k)^3 (N+3k)}}4}
\nn
\ea
and the shift of the $|\psi^k(l)|$ distribution from $l=N/4$, 
\bea
\mathfrak{S} &=& \frac{\omega_\perp}{m_0\Omega^2} \left[
						\frac{G_2 \omega_\perp}{\Lambda^2 N}
						\left(-\frac14 (k+N) + k\sqrt{\frac{N-k}{N+3 k}}\right) \right.\nn 
						& & \qquad\qquad \left. +\frac{1}{\Lambda^2} + \Lambda^2 
					\right].
\ea
				
Finally, the $l$ independent (but $k$ dependent) energy shift reads
\bea
E_0 &=& 
\frac{\omega_\perp}{4 \Lambda^2}
\left[
\frac{G_2}{8 N}\left(15 N^2 -2kN -5k^2\right) 
\right.
\nn
&& \left. +\left(1 + \Lambda^4\right) \left(3N+k\right)
\frac{}{}\!\!\right]
-\frac{1}{m_0} -\frac12 m_0\Omega^2\mathfrak{S}^2.
\ea
In these expressions, we have substituted the matrix elements $B_i$ with their values obtained from the variational ansatz in \eqref{B_i}, and the
corresponding single-particle energies $\epsilon_i$. 

The full ground state 
energy in this limit reads $E=\Omega/2 + E_0$ and its variational minimum at $\Lambda_{\rm 3m}$ can be solved for analytically. We do not display the 
very lengthy resulting expression here.
The Taylor expansion around $k/N=0$ reads, cf. the expression 
for the two-mode case in \eqref{2mLambda}, 
\bea
\Lambda_{\rm 3m}^4 &=& \Lambda_{\rm 2m}^4 + \frac{G_2}{12G_{>}}
		\left(448 - 151 G_2 + 56 G_{>}\right) \left(\frac kN\right)^2 
		\nn%
	& & +\ord\left(\frac{k}{N}\right)^3
\ea
with $G_{>}=\sqrt{16 - 80 G_2 + (G_2)^2}$. 
 
Exactly as in the quasi-2D two mode case, at the transition point where $G_{>}$ becomes real, given by ${(G_2)}_c$ in \eqref{G2crit2m},
the continuum limit becomes valid, and matches the newly appearing minimum of the numerical solution of the eigenvalue problem \eqref{FullEV2Dmod}. 
We then precisely recover the two-mode energy in the limit $N\to\infty$ for $k=0$.
This is a nontrivial result, because  
the corresponding states in the Fock basis do not coincide.
Examining the resulting energy minimum at $\Lambda_{\rm 3m}$, we observe a monotonous increase with the eigenvalue index $k$ and can hence
justify focussing only on minimal energy solutions for states around $k$. Below, we discuss the effects 
of quasi degenerate states for small $k$ whose occurrence can be understood in the continuum limit by showing that $\partial_k E_{\rm min}(k)|_{k=0}=0$.


For finite particle numbers, comparing the ground-state energies for the three- and two-mode ans\"atze, we find that the
three-mode model yields a slightly lower energy at $k=0$, of order $0.001 \omega_\perp$ per particle for particle numbers of order $N\sim 10^4$, 
further decreasing with increasing $N$ and asymptotically approaching zero
due to the identity of the energies in the continuum limit.


\vspace*{-2em}
\begin{center}
\begin{figure}[t]
\centering
\includegraphics[width=.47\textwidth]{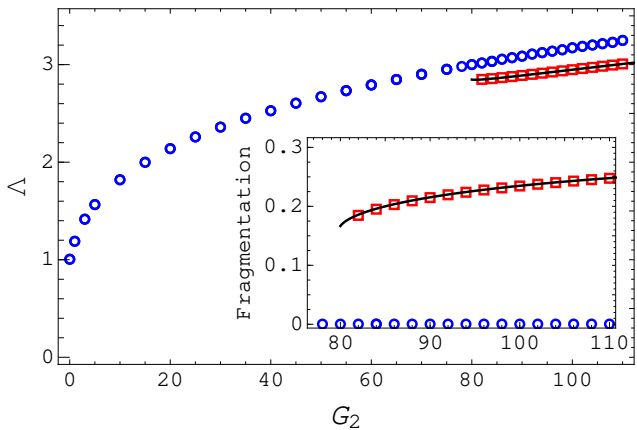}
\caption{\label{FragmentFig2D3mode} (Color online) 
Energy minima of the three-mode ansatz for the quasi-2D trapping geometry 
in terms of the variational parameter $\Lambda$, and three-mode degree of fragmentation $\mathfrak F$ (inset) from Eq.\,\eqref{F3m}, 
both as a function of the interaction measure $G_2$. 
Conventions are the same as those of Figs.\,\ref{FragmentFig1D} and \ref{FragmentFig2D}. 
The number of particles here is $N=20000$.}
\end{figure}
\end{center} 


The continuum limit, as well as numerical experiments for finite $N$ thus justify taking the minimal energy solution of the $k=0$ problem as 
the global ground state, which amounts to taking the state with equally populated counterclockwise and clockwise circulating single-particle modes.
However, we note that because of the similarity in structure for the eigenvalue problems of neighbouring values of $k$, the difference of the minimal energy eigenvalues per particle corresponding to different $k$ 
is of order $\omega_\perp/N$. For large particle numbers, we thus 
expect quasi-degeneracy of the eigenvalues located around $k=0$, because they
then  become arbitrarily closely spaced.

To study the consequences on fragmentation, we consider 
the one-particle density matrix 
whose off-diagonal elements are given by
\bea
\rho_{+-}
&=&\sum_{l,l'} \sqrt{l(l'+1)} \psi^*_{l,l'} \psi_{l-1,l'+1},\nn
\rho_{0+} &=& \sum_{l,l'} \sqrt{(N-l-l')(l+1)} \psi^*_{l,l'} \psi_{l+1,l'},
\label{nondiag}\\
\rho_{0-} &=& \sum_{l,l'} \sqrt{(N-l-l')(l'+1)} \psi^*_{l,l'} \psi_{l,l'+1},
\nonumber
\ea
and introduce a generalized fragmentation measure for three modes 
\bea
\mathfrak{F} = 1 - |\lambda_0-\lambda_2|/N, \label{F3m}
\ea 
where $\lambda_i$ are the eigenvalues of the one-particle density matrix ordered by size with $\lambda_0$ being the largest eigenvalue and $\lambda_2$ the smallest.

Supposing that the system only occupies a single state $k=0$, it is apparent from the $k$-diagonal form of \eqref{FullEV2Dmod} that all
off-diagonal elements of the density matrix in \eqref{nondiag} 
vanish. This is equivalent to a loss of coherence between states carrying different angular momenta.
In the continuum limit, 
the fragmentation can be written in terms of the shift as
$
\mathfrak{F} = 1 - \left|\frac14+\frac{\mathfrak{S}}{N/3}\right|
$
where the shift is defined by
$\langle a_0^\dagger a_0\rangle = N/2 + 2\mathfrak{S}$ with $-N/4\leq\mathfrak{S}\leq N/4$ (at $k=0$).
The fragmentation is then obtained to be, cf.\,Eq.\eqref{F2mcont},
\bea
 {\mathfrak F}(G_2) = 1- \left|\frac14+\frac{1-G_2/4+\Lambda_{\rm 2m}^4}{G_2}\right|  
\simeq \frac13 - \ord\left(\frac1{G_2}\right) 
.\nn \label{F3mcont} 
\ea
and has its asymptotic maximum value 33 \% at large $G_2 \gg {(G_2)}_c$, 
see also the inset of Fig.\,\ref{FragmentFig2D3mode}. 


In the quasi-degenerate case of large $N$ we expect many neighbouring low lying $k$ states to be occupied, leading for any small perturbation
to a mixing of different $k$ states away from $k=0$, and to nonvanishing off-diagonals \eqref{nondiag}.
Hereby coherence is established, and a nonfragmented ground state in terms of the {\em three-mode} single-particle density matrix and the corresponding fragmentation measure is obtained.
On the other hand, superposition of many modes of different $k$ leads to a (coherently superposed) excited state of zero angular momentum [taking into account both degenerate sectors corresponding to $l\le l'$ and $l\ge l'$], 
and we are then led back 
to two-mode fragmentation between ground state and the coherently
superposed excited state. This is due to the fact shown in section \ref{2msection}, 
namely that the two-mode fragmented state is always lower in energy than a single condensate above a critical $G_2$ given by \eqref{G2crit2m}.


\section{Discussion and Conclusions}
We considered harmonically trapped 
low-dimensional gases and found by a variational analysis that they split 
from a single macroscopically occupied field operator mode in the weakly confining
direction (the condensate) into two (quasi-1D trapping) respectively three (quasi-2D trapping) such macroscopically occupied modes with no remaining phase coherence between them, upon increasing a dimensionless measure of interaction strength beyond a critical value. We have furthermore demonstrated that 
due to the symmetry of the matrix equations for the state vector amplitudes,   three-mode fragmentation is highly susceptible to decay into a two-mode fragmented state in the  quasi-2D isotropic trapping geometry.
 
The results obtained represent, then, to the best of our knowledge 
the first example of ground-state fragmented scalar condensates in a single trap, and implement the 
result of \cite{Bader} in a physically realistic 
situation. They imply that the quasi-1D and quasi-2D condensates
decay into fragmented condensates of a few macroscopically occupied modes for a sufficiently large interaction energy, with the quasi-1D condensates, as could be expected, being significantly more fragile.   
By way of a numerical example for the quasi-1D situation, trapping frequencies of $\omega_z/2\pi = 3.5\,$Hz and $\omega_\perp/2\pi 
= 360\,$Hz (used in the magnetic trapping setup of \cite{Goerlitz}) 
correspond to a ratio $l_z/l_\perp \simeq 10$. For $^{23}$Na, 
we have $l_\perp \simeq 1\, \mu$m.  
For $a_s\simeq  2.8\,$nm, the number of atoms driving $G_1$
above its critical value is only of order $N_c \sim 500$. 
This raises the interesting question what was actually observed in the experiment \cite{Goerlitz}, 
where the quoted particle numbers are of order $N\sim 10^4$ for the quasi-1D setup.  
While the threshold to the first onset of fragmentation seems surprisingly low, 
one should note that the energy difference separating fragmented and nonfragmented condensate states is very small close to the threshold, Eq.\,\eqref{deltaElargeG1},  
the energy barriers between the two states being even smaller, cf.\,Fig.\,\ref{energyminima}. 
In order to observe the (zero temperature) 
transition point itself, extremely low temperatures would therefore be required.
Fragmented condensates become energetically clearly preferred 
over single condensates at experimentally  
accessible thermal fluctuation levels only at significantly higher values 
of the dimensionless interaction measures,
and therefore at much higher particle numbers for given trapping frequencies.
Another possibility to increase the interaction measures and 
to obtain fragmented condensate states 
is to increase the coupling constant itself \cite{Pollack}.


We have shown that fragmented condensate states in low-dimensional systems 
can be obtained well before the thermodynamic limit of infinite extension. While the degree of fragmentation depends on the (number of variational) modes chosen for a particular trapping setup as well as the applied fragmentation measure, the occurrence of few-mode condensate fragmentation for interacting bosonic gases should be a rather generic feature of their many-body physics. Few-mode fragmentation is owed to the confined nature of the system and a sufficiently strong and positive pair-exchange coupling between the bosonic modes. 

\acknowledgments
URF was supported by the Research Settlement Fund and College of Natural Sciences of Seoul National University, as well as the 
Basic Science Research Program of the National Research Foundation
of Korea (NRF), grant No. 2010-0013103. PB received support 
by the Generalitat Valenciana through the project GV/2009/032. In addition,
this research work was supported by the DFG under grant No. FI 690/3-1. 


\begin{thebibliography}{9999}
\bibitem{boundary} More precisely, first-order coherence vanishes up to boundary terms, i.e.,  
in the limit $N \rightarrow \infty$. We assume here that $A_3$ is real, so that the $\psi_l$ are
also real. Note that the thus defined first-order coherence function is zero even when a single condensate
exists, as the latter expresses nothing but the macroscopic population of the single $l=0$ mode (a single 
condensate is by definition a Fock state in $l$ space).
\bibitem{Corro} The assumption that at the nonfragmented local energy minimum to a good approximation 
all particles are residing in the energetically lower mode thus forming a single condensate,  
has been corroborated by the numerical analysis of \eqref{diffeq}
[which is possible as long as a nonfragmented minimum 
of the total energy exists, cf.\,Fig.\,\ref{energyminima}].
\end{thebibliography}
\end{document}